\documentclass{article}
\usepackage[psamsfonts]{amsfonts}
\usepackage[dvipsone]{epsfig}

\begin{document}
%%%%%%%%%%%%%%%%%%%%%%

\setcounter{page}{1}

\LRH{A. Cuzzocrea, J. Darmont, and H. Mahboubi}

\RRH{Fragmenting Very Large XML Data Warehouses via \emph{K}-Means
Clustering Algorithm}

\VOL{x}

\ISSUE{x}

\BottomCatch

\PAGES{xxxx}

\CLline

\PUBYEAR{2009}

\subtitle{}

\title{Fragmenting Very Large XML Data Warehouses via \emph{K}-Means Clustering Algorithm}

\authorA{Alfredo Cuzzocrea$^{1}$, J\'{e}r\^{o}me Darmont$^{2}$, and Hadj Mahboubi$^{2}$}
\affA{
$^{1}$ ICAR-CNR \& University of Calabria\\
Via P. Bucci, 41C, Rende\\
87036 Cosenza, Italy\\
E-mail: cuzzocrea@si.deis.unical.it\\
$^{2}$ University of Lyon (ERIC Lyon 2)\\
5 avenue Pierre Mend\`{e}s-France\\
69676 Bron Cedex, France\\
E-mail: \{hadj.mahboubi, jerome.darmont\}@eric.univ-lyon2.fr}

\begin{abstract}
XML data sources are more and more gaining popularity in the context of a wide family of \emph{Business Intelligence} (BI) and \emph{On-Line Analytical Processing} (OLAP) applications, due to the amenities of XML in representing and managing semi-structured and complex multidimensional data. As a consequence, many XML data warehouse models have been proposed during past years in order to handle heterogeneity and complexity of multidimensional data in a way traditional relational data warehouse approaches fail to
achieve. However, XML-native database systems currently suffer from limited performance, both in terms of volumes of manageable data and query response time. Therefore, recent research efforts are focusing the attention on \emph{fragmentation techniques}, which are able to overcome the limitations above. Derived horizontal fragmentation is
already used in relational data warehouses, and can definitely be adapted to the XML context. However, classical fragmentation algorithms are not suitable to control the number of originated fragments, which instead plays a critical role in data warehouses, and, with more emphasis, distributed data warehouse architectures. Inspired by this research challenge, in this paper we propose the use of \emph{K}-means clustering algorithm for effectively and efficiently supporting the fragmentation of very large XML data
warehouses, and, at the same time, completely controlling and determining the number of originated fragments via adequately setting the parameter $K$. We complete our analytical contribution by means of a comprehensive experimental assessment where we compare the efficiency of our proposed XML data warehouse fragmentation technique against those of classical derived horizontal fragmentation algorithms adapted to XML data warehouses.
\end{abstract}

\maketitle

%%%%%%%%%%%%%%%%%%%%%%%%%%%%%%%
\section{Introduction} %% 1
%%%%%%%%%%%%%%%%%%%%%%%%%%%%%%%

Nowadays, XML has become a standard for representing complex
business data~\cite{BeyerCCOP05}, so that decision support processes
that make use of XML data sources are now increasingly common.
However, XML data sources bear specificities that would be intricate
to handle in a relational environment. Among these specificities, we
recall: heterogeneous number and order of dimensions, complex
aggregation operations~\cite{ParkHS05} and measures, ragged
dimensional hierarchies~\cite{BeyerCCOP05}. Hence, many efforts
towards the so-called \emph{XML Data Warehousing} have been achieved
during the past few years~\cite{BoussaidMCA06,Pokorny02,ZhangWLZ05},
as well as efforts focused to extend \emph{XQuery}~\cite{xquery07}
with near \emph{On-Line Analytical Processing}
(OLAP)~\cite{OLAP93,GrayEtAl97} capabilities such as advanced
grouping and aggregation
features~\cite{BeyerCCOP05,datax08,WiwatwattanaJLS07}.

In this context, performance is a critical issue, as actual
XML-native database systems (e.g., \emph{eXist}~\cite{Meier02},
\emph{TIMBER}~\cite{timber02}, \emph{X-Hive}~\cite{X-Hive08}, and
\emph{Sedna}~\cite{sedna06}) suffer from limited performance, both
in terms of volumes of manageable manageable data and response time
to complex \emph{analytical queries}. These issues are well-known to
data warehouse researchers, and they can be addressed by means of
the so-called \emph{fragmentation techniques}~\cite{ZhangO94}.
Fragmentation consists in splitting a given data set into several
\emph{fragments} such that their combination yields the original
data warehouse without information loss nor information addition.
Fragmentation can subsequently support a meaningful
\emph{distribution} of the target data warehouse, e.g. on \emph{Data
Grids}~\cite{CostaF06} or across \emph{Peer-To-Peer} (P2P)
\emph{Networks}~\cite{KalnisNOPT02}. In the relational context,
\emph{derived horizontal fragmentation} is acknowledged as the
best-suited one to data warehouses~\cite{BellatrecheB05}. Basically,
this approach consists in fragmenting a given relation with respect
to \emph{query predicates} defined on another relation. Apart from
the above-mentioned research efforts, other XML data fragmentation
approaches have also been proposed
recently~\cite{BonifatiC07,BonifatiMCJ04,BoseF05,GertzB03,MaS03},
but they do not take into account multidimensional schemas
explicitly (i.e., star, snowflake, or fact constellation
schemas~\cite{kimball02}).

In derived horizontal fragmentation, dimensional tables first
undergo a \emph{primary horizontal fragmentation}. Output fragments
are then used to horizontally fragment the fact table into
sub-tables that each refer to a primary dimensional fragment. This
process is termed \emph{derivation}. Primary horizontal
fragmentation plays a critical role, as it heavily affects the
performance of the whole fragmentation process. In the relational
context, two major algorithms address this issue: \emph{predicate
construction}~\cite{NoamanB99} and
\emph{affinity-based}~\cite{NavatheKR95} algorithms. However, these
approaches suffer from an important limitation that makes them
unsuitable to XML Data Warehousing. In fact, in both algorithms the
number of fragments is not known in advance neither can be set as
input parameter, while in XML Data Warehousing it is crucial to
master this parameter, especially as distributing $M$ fragments over
$N$ nodes, with $M > N$, can be a critical issue in itself. In order
to become convinced of this aspect, it suffices to think of the
fragmentation problem in \emph{Distributed Data Warehousing}
environments~\cite{CostaF06}. Here, due to load-balancing and
scalability issues, node number can become very large, but
massive-in-size data warehouses can still represent a problematic
instance to be fragmented. Therefore, the need for completely
controlling the number of output fragments makes perfect sense.

Starting from these considerations, in this paper we propose
\emph{the usage of K-means~\cite{mcqueen67} clustering algorithm for
supporting the efficient fragmentation of XML data warehouses while
controlling the number of generated fragments through the parameter
$K$}. The latter specific feature has immediate benefits towards
efficiently supporting XML Data Warehousing in itself, as it will be
clear throughout the paper. Our proposed approach is inspired from a
proposal coming from the object-oriented databases
domain~\cite{Darabant04}. Summarizing, our proposal consists in
clustering the predicates of a reference query-workload posed to the
target XML data warehouse in order to produce primary horizontal
fragments from dimensional tables (XML documents, respectively),
with one fragment meaningfully corresponding to one \emph{cluster of
predicates}. Primary fragmentation is then derived on facts. Queries
based on predicates of the target query-workload are then evaluated
over the corresponding fragments only, instead of the whole data
warehouse, thus introducing a faster response time. The number of
fragments is directly related to the number of $K$-means-obtained
clusters (it is actually equal to $K+1$ --
Section~\ref{sec:FragmentConstruction}).

The remainder of this paper is organized as follows. In
Section~\ref{sec:RelatedWork}, we discuss state-of-the-art research
in fragmentation techniques for relational data warehouses and XML
databases, and also \emph{Data-Mining-based fragmentation
techniques}~\cite{Darabant04,FioletT05bis,GorlaJ05}, which, briefly,
propose applying Data Mining techniques in order to drive the
fragmentation phase. The latter is the class of techniques where our
research should be conceptually positioned.
Section~\ref{sec:XMLWarehouseReferenceModel} focuses the attention
on the XML data warehouse model we adopt as reference data model of
our research. In Section~\ref{sec:KMeansBasedFragmentation}, we
introduce our \emph{K}-means-based XML data warehouse fragmentation
approach. Section~\ref{sec:ExperimentalPerformanceStudy}
experimentally compares the efficiency of our proposed technique
against those of classical derived horizontal fragmentation
algorithms adapted to XML data warehouses, and shows its superiority
in accomplishing the desired goal. Finally,
Section~\ref{sec:Conclusion} contains conclusions of our research,
along with future research directions in fragmentation techniques
for XML data warehouses.

%%%%%%%%%%%%%%%%%%%%%%%%%%%%%%%
\section{Related Work} %% 2
\label{sec:RelatedWork}
%%%%%%%%%%%%%%%%%%%%%%%%%%%%%%%

In this Section, we first provide a brief taxonomy of relevant
fragmentation techniques, which have been originally proposed in the
relational context mainly. Then, we focus the attention on three
aspects that represent the conceptual/theoretical foundations of our
research, i.e. relational data warehouse fragmentation techniques,
XML database fragmentation techniques, and, finally,
Data-Mining-based fragmentation techniques.

\subsection{Taxonomy of Fragmentation Techniques}
\label{sec:taxo-fragmentation}

In the relational context, it is possible to identify three main
fragmentation techniques: \emph{vertical fragmentation},
\emph{horizontal fragmentation}, and \emph{hybrid fragmentation}.

Vertical fragmentation splits a given relation $R$ into
sub-relations that are \emph{projections} of $R$ with respect to a
subset of attributes. It consists in grouping together attributes
that are frequently accessed by queries. Vertical fragments are thus
built by projection. The original relation is reconstructed by
simply joining the fragments. Relevant examples for techniques
belonging to this class are the following. Navathe \textit{et al.}
vertically partition a relation into fragments and propose two
alternative fragmentation methods: \emph{progressive binary
partitioning}~\cite{Navathe84} and \emph{graphical
partitioning}~\cite{NavatheR89}. The first method is based on three
matrices (one capturing the \emph{Usage}, one capturing the
\emph{Affinity} and another one capturing the \emph{Coordinates} of
queries) while the second one exploits an objective function.
In~\cite{Navathe84}, authors present techniques for applying
vertical fragmentation in the following specialized application
contexts: databases stored on homogeneous devices, databases stored
in different memory levels, and distributed databases.

Horizontal fragmentation divides a given relation $R$ into sub-sets
of tuples by exploiting query predicates. It reduces query
processing costs by minimizing the number of irrelevant accessed
instances. Horizontal fragments are thus built by selection. The
original relation is reconstructed by fragment union. A variant, the
so-called derived horizontal fragmentation~\cite{BellatrecheB05},
consists in partitioning a relation $R$ with respect to predicates
defined on another relation, said $R'$. Other significant horizontal
fragmentation techniques are the following. Major algorithms that
address horizontal fragmentation are
\emph{Predicate-Construction-Based}~\cite{CeriP84} and the
\emph{Affinity-Based}~\cite{NavatheR89} methods
(Section~\ref{sec:relational-fragmentation}).

Finally, hybrid fragmentation consists of either horizontal
fragments that are subsequently vertically fragmented, or, by
contrary, vertical fragments that are subsequently horizontally
fragmented. Noticeable samples of these approaches are: (\emph{i})
\emph{Grid Creation}~\cite{NavatheKR95}, which proposes a mixed
fragmentation methodology allowing us to obtain a sub-optimal
partition of a given relation belonging to a distributed database,
and (\emph{ii}) \emph{View-Based Fragmentation}~\cite{PernulKN91},
which exploits views to build database fragments.

\subsection{Data Warehouse Fragmentation}
\label{sec:relational-fragmentation}

Several research studies address the issue of fragmenting relational
data warehouses, either to efficiently evaluate analytical queries,
or to efficiently distribute these data warehouses on settings like
data grids and P2P networks.

In order to improve ad-hoc query evaluation performance, Datta
\textit{et al.}~\cite{DattaRT99} propose exploiting a vertical
fragmentation of facts to build the index \emph{Cuio}, while
Golfarelli \textit{et al.}~\cite{GolfarelliMR99} propose applying
the same fragmentation methodology on data warehouse views. Munneke
\textit{et al.}~\cite{MunnekeWM99} instead propose an original
fragmentation methodology targeted to multidimensional databases. In
this case, fragmentation consists in deriving a global data cube
from fragments containing a sub-set of data defined by meaningful
slice and dice OLAP-like operations~\cite{OLAP93,GrayEtAl97}.
In~\cite{MunnekeWM99}, authors also define an alternative
fragmentation strategy, named \emph{server}, which removes one or
several dimensions from the target data cube in order to produce
fragments having fewer dimensions than the original data cube.

Bellatreche and Boukhalfa~\cite{BellatrecheB05} apply horizontal
fragmentation to data warehouse star schemas. Their fragmentation
strategy is based on a reference query-workload, and it exploits a
genetic algorithm to select a suitable partitioning schema among all
the possible ones. Overall, the proposed approach aims at selecting
an \emph{optimal fragmentation schema} that minimizes query cost. Wu
and Buchmaan~\cite{WuB97} recommend to combine horizontal and
vertical fragmentation for query optimization purposes.
In~\cite{WuB97}, a fact table can be horizontally partitioned with
respect to one or more dimensions of the data warehouse. Moreover,
the fact table can also be vertically partitioned according to its
dimensions, i.e. all the foreign keys to the dimensional tables are
partitioned as separate tables.

In order to distribute a data warehouse, Noaman \textit{et
al.}~\cite{NoamanB99} exploit a top-down strategy making use of
horizontal fragmentation. In~\cite{NoamanB99}, authors propose an
algorithm for deriving horizontal fragments from the fact table
based on input queries defined on all the dimensional tables.
Finally, Wehrle \textit{et al.}~\cite{WehrleMT05} propose
distributing and querying a data warehouse by meaningfully
exploiting the capabilities offered by a \emph{Computational Grid}.
In~\cite{WehrleMT05}, authors make use of derived horizontal
fragmentation to split the target data warehouse and build the
so-called \textit{block of chunks}, which is a set of data portions
derived from the data warehouse and used to query optimization
purposes, being each portion computed as a fragment of the
partition.

In summary, the above-outlined proposals generally exploit derived
horizontal fragmentation to reduce irrelevant data accesses and
efficiently process join operations across multiple relations
\cite{BellatrecheB05,NoamanB99,WehrleMT05}. From active
literature~\cite{KoreichiB97}, we also recognize that, in order to
implement derived horizontal fragmentation of data warehouses, the
outlined approaches prevalently make use of the following two main
fragmentation methods:

\begin{itemize}

\item \emph{Predicate-Construction-Based Fragmentation}~\cite{CeriP84} This method fragments a given relation
by using a complete and minimal set of predicates~\cite{NoamanB99}.
Completeness means that two relation instances belonging to the same
fragment have the same probability of being accessed by any
arbitrary query. Minimality guarantees that there is no redundancy
in predicates.

\item \emph{Affinity-Based Fragmentation}~\cite{NavatheR89} This method is an adaptation of
the vertical fragmentation approach~\cite{GolfarelliMR99} to the
horizontal fragmentation one~\cite{NavatheKR95}. It is based on the
\emph{predicate affinity concept}~\cite{ZhangO94} according to which
affinity is defined in terms of query frequency. Specific
predicate-usage and affinity matrices are exploited in order to
cluster selection predicates. A cluster is here defined as a
\emph{selection predicate cycle}, and forms a fragment of a
dimensional table itself.

\end{itemize}

\subsection{XML Database Fragmentation}
\label{sec:xml-fragmentation}

Recently, several fragmentation techniques for XML data have been
proposed in literature. These techniques propose splitting an XML
document into a new set of XML documents, with the main goal of
either improving XML query
performance~\cite{BonifatiC07,GertzB03,MaSHK03}, or distributing or
exchanging XML data over a network~\cite{BonifatiMCJ04,BoseF05}.

In order to fragment XML documents, Ma \textit{et
al.}~\cite{MaS03,MaSHK03} define a new fragmentation notion, called
\textit{split}, which is inspired from the oriented-object databases
context. This fragmentation technique splits elements of the input
XML document, and assigns a reference to each so-obtained
sub-element. References are then added to the \emph{Document Type
Definition} (DTD) defining the input XML document. This avoid
redundancy and inconsistence problems that could occur due to
fragmentation process. Bonifati \textit{et
al.}~\cite{BonifatiC07,BonifatiCZ06} propose a fragmentation
strategy for XML documents that is driven by the so-called
\emph{structural constraints}. These constraints refer to intrinsic
properties of XML trees such as the depth and the width of trees. In
order to efficiently fragment the input XML document by means of
structural constraint, the proposed strategy exploits heuristics and
statistics simultaneously.

Andrade \textit{et al.}~\cite{AndradeRBBM06} propose applying
fragmentation to an \emph{homogeneous} collection of XML documents.
In~\cite{AndradeRBBM06}, authors adapt traditional fragmentation
techniques to an XML document collection, and make use of the
\emph{Tree Logical Class} (TLC) algebra~\cite{PaparizosWLJ04} to
this goal. Authors also experimentally evaluate these techniques and
show that horizontal fragmentation provides the best performance.
Gertz and Bremer~\cite{GertzB03} introduce a distribution approach
for XML repositories. They propose a fragmentation method and
outline an allocation model for distributed XML fragments in a
centralized architecture. In~\cite{GertzB03}, authors also define
horizontal and vertical fragmentation for XML repositories. Here,
fragments are defined on the basis of a \emph{path expression
language}, called \textit{XF}, which is derived from
\emph{XPath}~\cite{xpath99}. In more detail, fragments are obtained
via applying an \textit{XF} expression on a graph representing XML
data, named \emph{Repository Guide} (\textit{RG}). Moreover, authors
provide exclusion expressions that ensure fragment coherence and
disjunction rigorously.

Bose and Fegaras~\cite{BoseF05}, argue to use XML fragments for
efficiently supporting data exchange in P2P networks. In this
proposal, XML fragments are interrelated, and each fragment is
univocally identified by an \textit{ID}. Authors also propose a
fragmentation schema, called \textit{Tag Structure}, which allows us
to define the structure of fragments across the network. In turn,
the structure of fragments can be exploited for data exchange and
query optimization purposes. Bonifati \textit{et
al.}~\cite{BonifatiMCJ04} also define an XML fragmentation framework
for P2P networks, called \emph{XPath-To-Partition} (XP2P). In this
proposal, XML fragments are obtained and identified via a single
root-to-node path expression, and managed on a specific peer. In
addition, to data management efficiency purposes,
in~\cite{BonifatiMCJ04} authors associate two XPath-modeled path
expressions to each fragment, namely \textit{super fragment} and
\textit{child fragment}, respectively. Given an XML fragment $f$,
the first XPath expression identifies the root of the fragment $f'$
from which $f$ has been originated; the second XPath expression
instead identifies the root of a $f$'s child XML fragment. These
path expressions ensure the easily identification of fragments and
their networked relationships.

In summary, the above-outlined proposals adapt classical
fragmentation methods, mainly investigated and developed in the
context of relation data warehouses, in order to split a given XML
database into a meaningfully collection of XML fragments. An XML
fragment is defined and identified by a path
expression~\cite{BonifatiMCJ04,GertzB03}, or an XML algebra
operator~\cite{AndradeRBBM06}. Fragmentation is performed on a
single XML document~\cite{MaS03,MaSHK03}, or an homogeneous XML
document collection~\cite{AndradeRBBM06}. Another secondary result
deriving from this is represented by the claim stating that, to the
best of our knowledge, XML data warehouse fragmentation has not been
addressed at now by active literature. This further confirms the
innovation carried out by our research.

\subsection{Data-Mining-based Fragmentation}
\label{sec:DataMiningBasedFragmentation}

Although Data Mining has already proved to be extremely useful to
select physical data structures that enhance performance, such as
indexes or materialized
views~\cite{agr00aut,adbis06,innovations07,zaman04}, few
fragmentation approaches that exploit Data Mining exist in
literature. Therefore, it is reasonable to claim that the latter is
a relatively-novel area of research, and a promising direction for
future efforts in data warehouse and database fragmentation
techniques.

Gorla and Betty~\cite{GorlaJ05} exploit \emph{association rules} for
vertical fragmentation of relational databases. Authors consider
that association rules provide a natural way to represent
relationships between attributes as implied by database queries.
Basically, their solution consists in adapting the well-known
algorithm Apriori~\cite{AgrawalS94} by selecting the non-overlapping
item-sets having highest support and by grouping their respective
attributes into one partition. Then, the algorithm exploits a cost
model to select an optimal fragmentation schema.Darabant and
Campan~\cite{Darabant04} propose using \emph{K}-means clustering for
efficiently supporting horizontal fragmentation of object-oriented
distributed databases. This research has inspired our work. In more
detail, the method proposed in~\cite{Darabant04} clusters object
instances into fragments via taking into account all complex
relationships between classes of data objects (aggregation,
associations and links induced by complex methods). Finally, Fiolet
and Toursel~\cite{FioletT05bis} propose a parallel, progressive
\emph{clustering algorithm} to fragment a database and distribute it
over a data grid. This approach is inspired by the sequential
clustering algorithm \emph{CLIQUE}~\cite{Agrawal98} that consists in
clustering data by means of projection operations.

Even though in limited number, these studies clearly demonstrate how
Data Mining can be efficiently used to support horizontal and
vertical fragmentation of both data warehouses and databases,
throughout association rule mining and clustering, respectively.

%%%%%%%%%%%%%%%%%%%%%%%%%%%%%%%%%%%%%%%%%%%%%%%%%%%%%%%
\section{A Reference XML Data Warehouse Model} %% 3
\label{sec:XMLWarehouseReferenceModel}
%%%%%%%%%%%%%%%%%%%%%%%%%%%%%%%%%%%%%%%%%%%%%%%%%%%%%%%

Actual XML data warehouse models from the
literature~\cite{GolfarelliRV01,ParkHS05,Pokorny02} share a lot of
concepts, mostly originating from classical results developed in the
relational context. Despite this common origin, actual XML data
warehouse models are nonetheless all different. From this evidence,
in~\cite{edwm208} a unified, reference XML data warehouse model that
synthesizes and enhances existing models is proposed. This proposal
represents the fundamental data model of our proposed XML data
warehouse fragmentation technique. Given this significant
relationship between~\cite{edwm208} and the the research we propose
in this paper, before to detail our XML data warehouse fragmentation
approach (Section~\ref{sec:KMeansBasedFragmentation}), in this
Section we review the XML data warehouse model~\cite{edwm208}.

State-of-the-art XML data warehouse models assume that the target
data warehouse is composed by XML documents representing both facts
and dimensions. All these studies mostly differ in the way
dimensions are handled, and the number of XML documents that are
used to store facts and dimensions. A performance evaluation study
of these different representations has shown that representing facts
in one singleton XML document and each dimension in one singleton
XML document allows the best performance~\cite{DoulkiliBB06}.
Moreover, the above representation model also allows us to model
fact constellation schemas without the need of duplicating dimension
information, thus achieving the so-called \emph{shared
dimensions}~\cite{kimball02}. This has several benefits for what
concerns with the scalability of the model, which is an extremely
critical factor in Data Warehousing. According to this
representation model, several fact documents can indeed share the
same dimensions. Hence, we adopt this architecture model. In more
detail, our reference XML data warehouse model is composed by the
following XML documents:

\begin{itemize}
\item $dw-model.xml$, which stores warehouse metadata;
\item a set of documents $facts_f.xml$, such that each document stores information related
to a set of facts $f$;
\item a set of documents $dimension_{d}.xml$, such that each documents stores the member values of
the dimension $d$.
\end{itemize}

\begin{figure}[hbt]
\centering \epsfig{file=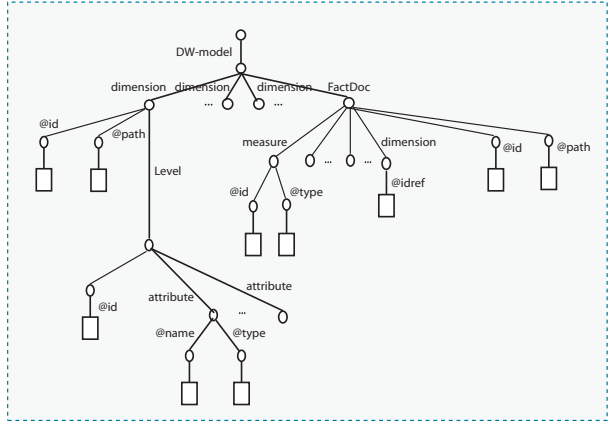, width=8cm} \caption{The XML
Document $dw-model.xml$} \label{fig:dw-model}
\end{figure}

Document $dw-model.xml$ (Figure~\ref{fig:dw-model}) defines the
multidimensional structure of the target data warehouse. The root
node, named as \verb"DW-model", is composed by two kinds of nodes:
\verb"dimension" and \verb"FactDoc", respectively.

A \verb"dimension" node models a dimension of the data warehouse. In
a \verb"dimension" node, the following elements are contained:
(\emph{i}) element \verb"@id" that models the absolute identifier of
the dimension $d$; (\emph{ii}) element \verb"@path" that models the
path to the corresponding document $dimension_{d}.xml$ storing the
related dimension information; (\emph{iii}) a set of \verb"Level"
elements, such that each element models a level $L$ of the
(possible) hierarchical levels of the dimension $d$. Under a
\verb"Level" element, we have: (\emph{i}) element \verb"@id" that
models the absolute identifier of the level $L$; (\emph{ii}) a set
of \verb"attribute" elements, such that each element models an
attribute $a$ of the level $L$. Under an \verb"attribute" element,
we have: (\emph{i}) element \verb"@name" that models the name of the
attribute $a$; (\emph{ii}) element \verb"@type" that models the type
of the attribute $a$.

A \verb"FactDoc" node models a fact of the data warehouse. In a
\verb"FactDoc" node, the following elements are contained:
(\emph{i}) element \verb"@id" that models the absolute identifier of
the fact $f$; (\emph{ii}) element \verb"@path" that models the path
to the corresponding document $facts_{f}.xml$ storing the related
fact information; (\emph{iii}) a set of elements \verb"measure",
such that each element models a measure $m$ of the fact $f$;
(\emph{iv}) a set of \verb"dimension" elements, such that each
element references a dimension $d$ of the XML data warehouse schema.
Under a \verb"measure" element, we have: (\emph{i}) element
\verb"@id" that models the absolute identifier of the measure $m$;
(\emph{ii}) element \verb"@type" that models the type of the measure
$m$. Under a \verb"dimension" element, we have the element
\verb"@idref" that models the reference to the corresponding
dimension $d$ of the XML data warehouse schema.

\begin{figure}[hbt]
\centering \epsfig{file=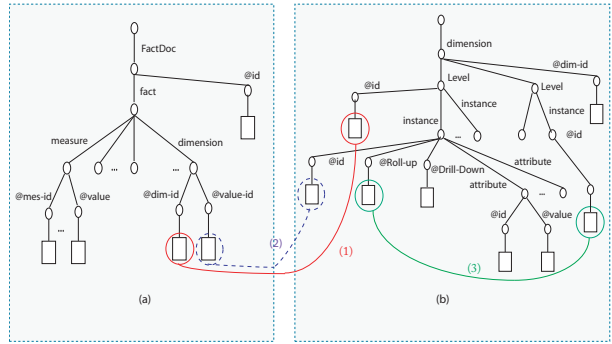, width=8cm}
\caption{The XML Documents $facts_{f}.xml$ (a) and
$dimension_{d}.xml$ (b)} \label{fig:fact-dimension}
\end{figure}

Figure~\ref{fig:fact-dimension} shows the structure of a document
$facts_{f}.xml$ (Figure~\ref{fig:fact-dimension}(a)) and a document
$dimension_{d}.xml$ (Figure~\ref{fig:fact-dimension}(b)),
respectively. The Figure also details the relationship between facts
and dimensions, and how this relationship is captured in our
reference XML data warehouse model. A $facts_{f}.xml$ document
(Figure~\ref{fig:fact-dimension}(a)) stores facts. It is structured
in terms of the document root node, \verb"FactDoc", which contains
an element \verb"@id" that models the absolute identifier of the
fact, and a set of elements \verb"fact", such that each element
instantiates a fact of the XML data warehouse schema in terms of
measure values and dimension references. Here, measures and
dimensions are modeled in a similar way to what provided for the
document $dw-model.xml$ storing the warehouse metadata. The
fact-to-dimension relationship is captured by means of conventional
XML identifiers. Finally, a $dimension_{d}.xml$ document
(Figure~\ref{fig:fact-dimension}(b)) stores a dimension, including
its possible hierarchical levels. The document root node,
\verb"dimension", contain the following nodes: (\emph{i}) element
\verb"@dim-id" that models the absolute identifier of the dimension;
(\emph{ii}) a set of elements \verb"Levels", such that each element
models a level $L$ of the dimension $d$, and contains a collection
of elements \verb"instance" that defines member attribute values $v$
of the level $L$. Overall, this allows us to model an OLAP
hierarchical level in all its characteristics and values. Here,
attributes are modeled in a similar way to what provided for the
document $dw-model.xml$ storing the warehouse metadata. In addition,
an \verb"instance" element also contains the elements
\verb"@Roll-Up" and \verb"@Drill-Down", respectively, which both
define the hierarchical relationship of the actual level within the
modeled dimension, and support classical OLAP data cube exploration
operations.

\subsection{Example}
\label{sec:DW-Example}

\begin{figure}[hbt]
\centering \epsfig{file=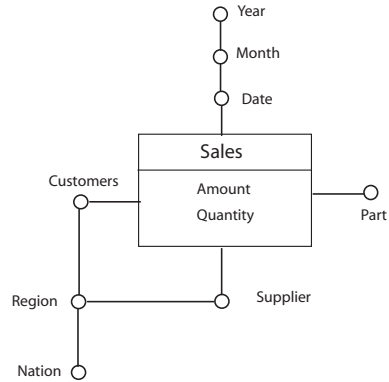, width=5cm} \caption{The Sample Data
Warehouse \emph{Sales}} \label{fig:dfm}
\end{figure}

\begin{figure}[hbt]
\centering \epsfig{file=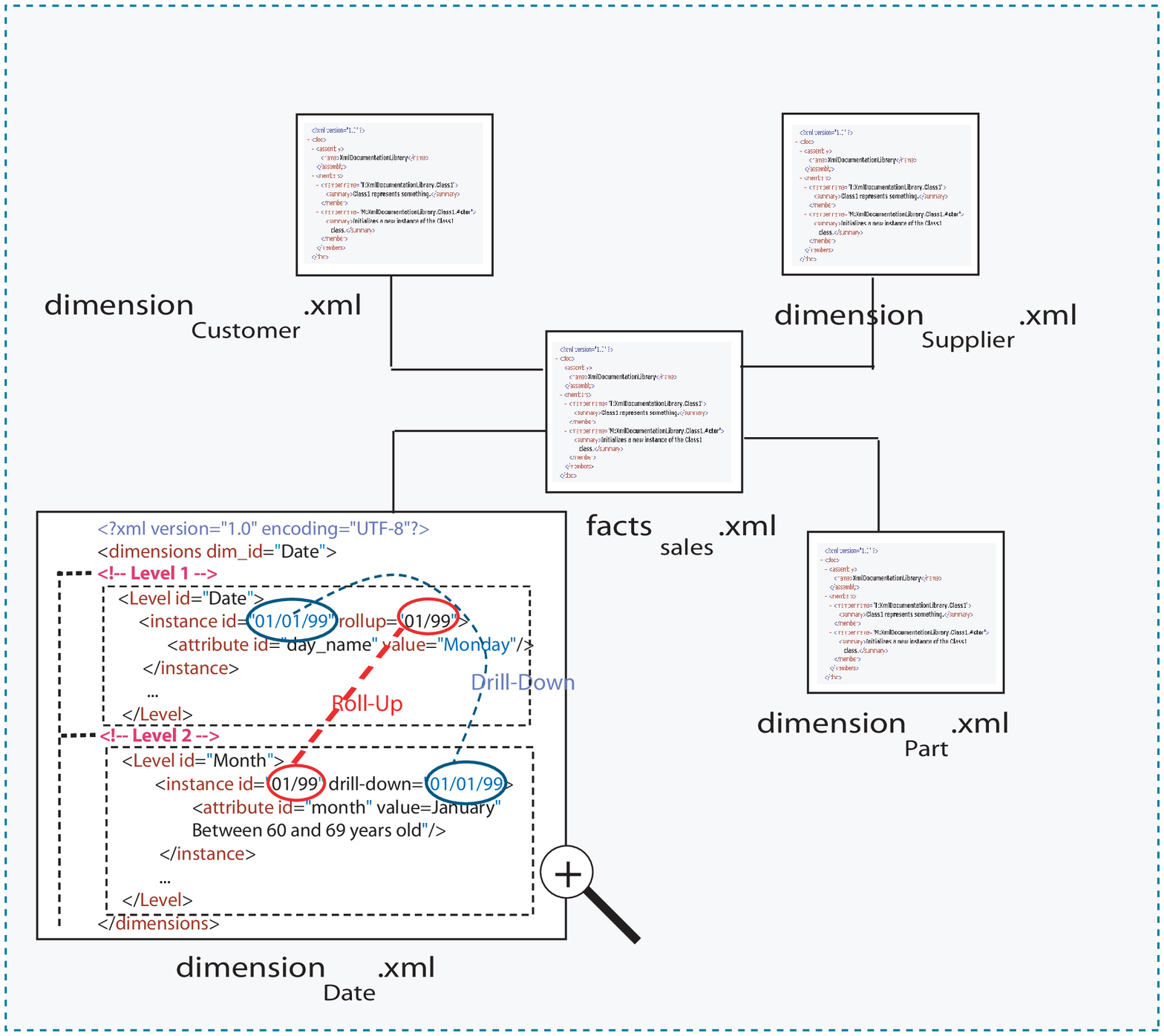, width=9cm} \caption{XML Documents
associated to the Data Warehouse \emph{Sales} of
Figure~\ref{fig:dfm}} \label{fig:model}
\end{figure}

In this Section, we provide a sample four-dimensional XML data
warehouse represented by means of our reference data model. Consider
the \emph{Dimensional Fact Model} (DFM)~\cite{GolfarelliMR98}
depicted in Figure~\ref{fig:dfm}, which models the data warehouse
\emph{Sales} one can find in a typical retail application. In this
schema, \emph{Quantity} and \emph{Amount} play the roles of measure,
whereas \emph{Customer}, \emph{Supplier}, \emph{Part} and
\emph{Date} play the roles of dimension. Figure~\ref{fig:model}
provides instead an overview of the set of XML documents that,
according to our reference model, describes the data warehouse
\emph{Sales}.

%%%%%%%%%%%%%%%%%%%%%%%%%%%%%%%%%%%%%%%%%%%%%%%%%%%%%%%%%%%%%%%%%%%%%%%%%%%%
\section{\emph{K}-Means-based Fragmentation of XML Data Warehouses} %% 4
\label{sec:KMeansBasedFragmentation}
%%%%%%%%%%%%%%%%%%%%%%%%%%%%%%%%%%%%%%%%%%%%%%%%%%%%%%%%%%%%%%%%%%%%%%%%%%%%

In this Section, we present and discuss our \emph{K}-means-based
fragmentation approach for XML data warehouses. In this respect, we
first provide an overview on the proposed technique, by highlighting
the fundamental tasks it is composed, and then we focus the
attention on each of these tasks in a greater detail.

\subsection{Overview}
\label{sec:Principle}

Since the aim of fragmentation is that of optimizing query response
time, the prevalent fragmentation strategies are
workload-driven~\cite{BellatrecheB05,BonifatiC07,GertzB03,NavatheKR95,NoamanB99},
i.e. they assume a reference query-workload and try to optimize
queries belonging to this query-workload rather than any arbitrary
query than can be posed to the target data warehouse. We highlight
that, fixing a reference query-workload $QW$, does not mean to
efficiently answer queries in $QW$ solely and discard the other
(still possible) queries, but rather that queries in $QW$ represent
a set of queries that (\emph{i}) are \emph{probabilistically} likely
to be posed to the data warehouse, and (\emph{ii}) any other
arbitrary query to the data warehouse is \emph{probabilistically}
likely to be ``similar'' to queries in $QW$. Therefore, the final
goal is that of exploiting query-workload information to improve
query evaluation. For what regards practical issues, it should be
noted that any conventional \emph{Data Warehouse Server} embeds
monitoring tools that are able to gather statistics on the query
flow posed to the server. These statistics, which are originally
meant for data warehouse maintenance and tuning (e.g., index
tuning), represent an invaluable source of information to define and
model query-workloads, even complex in nature (e.g., analytical
queries).

\begin{figure}[hbt]
\centering \epsfig{file=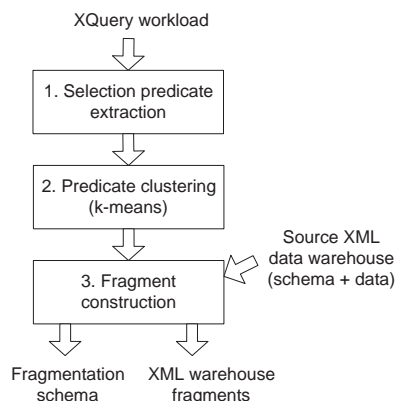, width=6cm}
\caption{\emph{K}-Means-Based XML Data Warehouse Fragmentation
Overview} \label{fig:principle}
\end{figure}

The approach used to effectively exploit the information embedded
into the query-workload can be exploited in different ways,
depending on the particular application scenario considered (e.g.,
relational databases, peer-to-peer databases, object-oriented
databases, and so forth). In the particular context represented by
the fragmentation of data warehouses, state-of-the-art approaches
exploit \emph{selection predicates} of workload queries in order to
derive \emph{suitable fragments}. Our proposed approach still
belongs to this family. Figure~\ref{fig:principle} sketches our
\emph{K}-means-based XML data warehouse fragmentation technique. The
proposed technique takes as input the XML data warehouse (both
including schema and instance) and the reference query-workload. It
returns as output the fragmented XML warehouse and the so-called
\emph{fragmentation schema}, which are ad-hoc meta-data describing
how the data warehouse has been fragmented and schemas of fragments.
These schemas are definitively useful to query optimization
purposes. As intermediate steps, the following ones arise:
(\emph{i}) extraction of selection predicates from the workload
queries; (\emph{ii}) predicate clustering by means of algorithm
\emph{K}-means; (\emph{iii}) fragment construction with respect to
predicate clusters generated at the previous step.

\subsection{Extraction of Selection Predicates}
\label{sec:SelectionPredicateExtraction}

Given a query-workload $QW$, the output selection predicate set $SP$
is obtained by simply parsing queries in $QW$ and extracting the
predicates of such queries. For instance, consider
Figure~\ref{fig:req}, where a sample XQuery workload $QW = \{q_{1},
q_{2}, ..., q_{10}\}$ is depicted. Figure~\ref{fig:pred} shows
instead a portion of the output selection predicate set $SP =
\{p_{1}, p_{2}, p_{3}, p_{4}, ...\}$, which has been geenrated
according to our proposed approach. Here, for instance, $p_2$ and
$p_3$ are selection predicates obtained from query $q_{2} \in QW$.
It should be noted how actually a large number of XML parsing tools
such as \emph{Java DOM}~\cite{jdom09} are available in order to
adequately fulfill the application requirement determined by the
selection predicate extraction phase of our proposed XML data
warehouse fragmentation technique.

\begin{figure}[hbt]
\centering{ \scriptsize{
\begin{tabular}{cl}
$q_{1}$ &   for \$x in //FactDoc/Fact,\\
&           \$y in //dimension[@dim-id="Customer"]/Level/instance\\
&            where \$y/attribute[@id="c\_nation\_key"]/@value$>$"15"\\
&          and \$x/dimension[@dim-id="Customer"]/@value-id=\$y/@id\\
&       return \$x\\ \\
$q_{2}$ &   for \$x in //FactDoc/Fact,\\
&           \$y in //dimension[@dim-id="Customer"]/Level/instance,\\
&           \$z in //dimension[@dim-id="Part"]/Level/instance\\
&            where \$y/attribute[@id="c\_nation\_key"]/@value="13"\\
&            and \$y/attribute[@id="p\_type"]/@value="PBC"\\
&          and \$x/dimension[@dim-id="Customer"]/@value-id=\$y/@id\\
&          and \$x/dimension[@dim-id="Part"]/@value-id=\$z/@id\\
&       return \$x\\
\dots \\
$q_{10}$ &   for \$x in //FactDoc/Fact,\\
&           \$y in //dimension[@dim-id="Customer"]/Level/instance,\\
&           \$z in //dimension[@dim-id="Date"]/Level/instance\\
&            where \$y/attribute[@id="c\_nation\_key"]/@value="13"\\
&            and \$y/attribute[@id="d\_date\_name"]/@value="Sat"\\
&          and \$x/dimension[@dim-id="Customer"]/@value-id=\$y/@id\\
&          and \$x/dimension[@dim-id="Part"]/@value-id=\$z/@id\\
&       return \$x\\
\end{tabular}}
} \caption{A Sample XQuery Workload}\label{fig:req}
\end{figure}

\begin{figure}[hbt]
\centering{ \scriptsize{
\begin{tabular}{l}
$p_1$ = \$y/attribute[@id="c\_nation\_key"]/@value$>$"15"\\
$p_2$ = \$y/attribute[@id="c\_nation\_key"]/@value="13"\\
$p_3$ = \$y/attribute[@id="p\_type"]/@value="PBC"\\
$p_4$ = \$y/attribute[@id="d\_date\_name"]/@value="Sat"
\end{tabular}}
} \caption{Some Selection Predicates Extracted from the Sample
XQuery Workload of Figure~\ref{fig:req}}\label{fig:pred}
\end{figure}

Parsed predicates are then coded into a \emph{Query-Predicate
Matrix} $QP$, whose general term $qp_{ij}$ is equal to $1$ if the
predicate $p_{j} \in SP$ appears in the query $q_{i} \in QW$,
otherwise it is equal to $0$. For instance, the matrix $QP$ derived
from the query-workload $QW$ of Figure~\ref{fig:req} and the
selection predicate set $SP$ of Figure~\ref{fig:pred} is featured in
Table~\ref{Table:qps}.

\begin{table}[hbt]
\begin{center}
\begin{tabular}{|l|c|c|c|c|c|}
\hline
& \small{$p_1$} & \small{$p_2$} & \small{$p_3$} & \small{$p_4$} &  \small{...} \\
\hline
\small{$q_1$} & \small{1} & \small{0} & \small{0} & \small{0} & \\
\hline
\small{$q_2$} & \small{0} & \small{1} & \small{1} & \small{0} & \\
\hline
\small{...} & & & & & \\
\hline
\small{$q_{10}$} & \small{0} & \small{0} & \small{1} & \small{1} & \\
\hline
\end{tabular}
\caption{The Query-Predicate Matrix $QP$ derived from the
Query-Workload of Figure~\ref{fig:req} and the Selection Predicate
Set of Figure~\ref{fig:pred}} \label{Table:qps}
\end{center}
\end{table}

It should be noted that, being matrix-based, the proposed approach
could expose scalability issues. In particular, these problem could
occur in the presence of query-workloads characterized by a high
cardinality, and too ``dense'' queries, i.e. queries defined on top
of a significant number of predicates. In turn, this originates a
large number of rows and a large number of columns in the
Query-Predicate matrix, respectively. In more detail, the number of
columns of the Query-Predicate matrix also depends on the degree of
similarity/dissimilarity between selection predicates embedded in
the target query-workload. Similarly, the opposite problem could be
experienced. When query-workloads characterized by a low cardinality
and too ``sparse'' queries, i.e. queries defined on top of a small
number of predicates, are handled, the extracted information (i.e.,
the selection predicate set) could not be enough to fulfill the goal
of building a ``reliable'' input for algorithm \emph{K}-means.
Contrary to the previous case, in this special case the derived
Query-Predicate matrix is sparse. While both topics are very
interesting and should merit a proper research effort, they are
outside the scope of this paper, and we will hereafter assume of
dealing with query-workloads that do not expose ``problematic''
characteristics whose some instances have been mentioned above.

\subsection{Predicate Clustering}
\label{sec:SelectionPredicateClustering}

The main goal of our XML data warehouse fragmentation technique
consists in obtaining fragments able to optimize data accesses for
queries of the target query-workload. In turn, this allows us to
take advantages in the query evaluation phase, as the overall
response time of typical queries posed to the data warehouse (e.g.,
OLAP queries) can be lowered. Since horizontal fragments
(Section~\ref{sec:taxo-fragmentation}) are built from selection
predicates, clustering predicates with respect to queries achieves
the goal above. Predicates that are syntactically similar are indeed
grouped in a same cluster, which helps building an horizontal
fragment. Intuitively enough, we ideally aim at building rectangles
of 1s in the Query-Predicate matrix $QP$ that correspond to clusters
of predicates, as 1 denotes the occurrence of a predicate $p_j$ in a
certain query $q_i$. To this end, in our proposal we adopt the
widely-used clustering algorithm \emph{K}-means in order to
effectively accomplish this task.

Given a data set $D$, algorithm \emph{K}-means takes as input a
vector of object attributes of $D$ (i.e., predicates as columns of
the Query-Predicate matrix $QP$, in our case), and returns as output
a set of $K$ clusters $C = \{C_{1}, C_{2}, ..., C_{K}\}$ by finding
the \emph{centers} of so-called ``natural''
clusters~\cite{Carmichael68} in $D$ via minimizing the total
\emph{intra-cluster variance} of $C$, which is defined as follows:

\begin{center}
$\sum_{i=1}^k \sum_{x_j \in C_i} (x_j - \mu_i)^2$
\end{center}

where $x_j$ denotes a data item in $D$ belonging to a certain
cluster $C_i \in C$, and $\mu_i$ denotes the \emph{centroid} (i.e.,
the mean point) of data items $x_j \in C_i$.

Usually, having $K$ as an input parameter is viewed as a drawback
for clustering algorithms, as this limits the quality of the final
cluster set obtained. Contrary to this, in our proposed XML data
warehouse fragmentation technique this peculiarity turns to be an
advantage, since we aim at controlling and limiting the number of
clusters/fragments generated by the fragmentation approach. As a
baseline guideline, $K$ could be set as equal to the number of nodes
the XML data warehouse will be distributed on.

In our fragmentation framework, in order to exploit a reliable
already-available implementation of \emph{K}-means, we make use of
\emph{Weka}~\cite{weka}, a collection of Machine Learning algorithms
for Data Mining tasks. In more detail, we exploit the Weka's
\textsf{SimpleKMeans} implementation of \emph{K}-means. Rather than
more complex ones, \textsf{SimpleKMeans} makes use the
\emph{Euclidean} distance for computing distances between data items
and clusters. Looking at our specific case, \textsf{SimpleKMeans}
takes as input the matrix $QP$ (actually, the vector of predicates
$p_j \in SP$) and the parameter $K$, and returns as output the set
of predicate clusters $C$. For instance, consider the
Query-Predicate matrix $QP$ of Table~\ref{Table:qps}. By setting $K
= 2$, \textsf{SimpleKMeans} produces the following output:

\begin{center}
$C = \{\{p_1\}, \{p_2, p_3, p_4\}\}$
\end{center}

Finally, it should be noted how our proposed XML data warehouse
fragmentation framework is indeed open to be customized for any
other clustering algorithm beyond \emph{K}-means. This nice feature,
which makes our framework orthogonal to the particular clustering
algorithm chosen, is indeed due to the independence ensured by the
Query-Predicate matrix, on which any clustering algorithm can run.

\subsection{Fragment Construction}
\label{sec:FragmentConstruction}

\begin{figure}[hbt]
\centering \epsfig{file=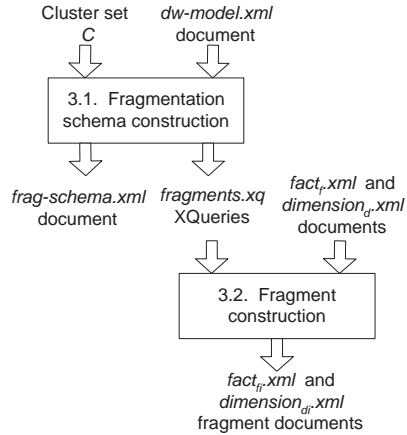, width=6.1cm}
\caption{Fragment Construction Sub-Steps} \label{fig:fragments}
\end{figure}

\begin{figure}[hbt]
\centering \epsfig{file=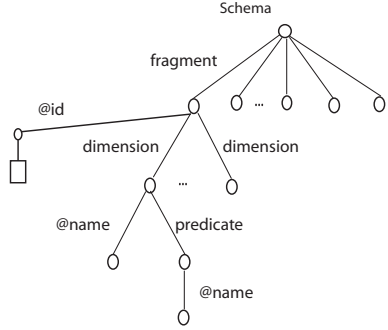, width=5cm}
\caption{The XML Document $frag-schema.xml$}
\label{fig:frag-schema}
\end{figure}

The fragmentation construction step of our XML data warehouse
fragmentation technique is composed by two sub-steps
(Figure~\ref{fig:fragments}), the fragment schema construction and
the proper fragment construction, respectively. In the first step,
predicate cluster set $C$ is joined to the warehouse schema stored
in the document $dw-model.xml$ in order to produce a new XML
document named as $frag-schema.xml$ that models the fragmentation
schema (Figure~\ref{fig:frag-schema}). The root node of
$frag-schema.xml$, called \verb"Schema", is composed by a set of
\verb"fragment" elements. Each \verb"fragment" element models a
fragment $f$ generated by the fragmentation process. A
\verb"fragment" element contains the element \verb"@id", which
models the absolute identifier of the fragment $f$, and a set of
elements \verb"dimension", which model the warehouse dimensions. A
\verb"dimension" element contains the element \verb"@name", which
models the name of the dimension $d$, and the element
\verb"predicate", which stores the predicate $p$ used for the
fragmentation process. Finally, a \verb"predicate" element contains
the element \verb"@name", which models the name of the predicate
$p$. To give an example, consider Figure~\ref{fig:frag-doc}, where
the fragmentation schema corresponding to the cluster set $C$ of the
running example (Section~\ref{sec:SelectionPredicateClustering}) is
shown.

\begin{figure}[hbt]
\centering{ \scriptsize{
\begin{tabular}{p{6.5cm}}
$<$Schema$>$\\
    \hspace*{1cm} $<$fragment id="f1"$>$\\
        \hspace*{2cm} $<$dimension name="Customer"$>$\\
            \hspace*{3cm} $<$predicate name="p1"/$>$\\
        \hspace*{2cm} $<$/dimension$>$\\
    \hspace*{1cm} $<$/fragment$>$\\
    \hspace*{1cm} $<$fragment id="f2"$>$\\
        \hspace*{2cm} $<$dimension name="Customer"$>$\\
            \hspace*{3cm} $<$predicate name="p2"/$>$\\
        \hspace*{2cm} $<$/dimension$>$\\
        \hspace*{2cm} $<$dimension name="Part"$>$\\
            \hspace*{3cm} $<$predicate name="p3"/$>$\\
        \hspace*{2cm} $<$/dimension$>$\\
        \hspace*{2cm} $<$dimension name="Date"$>$\\
            \hspace*{3cm} $<$predicate name="p4"/$>$\\
        \hspace*{2cm} $<$/dimension$>$\\
    \hspace*{1cm} $<$/fragment$>$\\
$<$/Schema$>$
\end{tabular}}
} \caption{The Output XML Document $frag-schema.xml$ corresponding
to the Running Example Fragmentation Process} \label{fig:frag-doc}
\end{figure}

The fragment schema construction sub-step also outputs a set of
XQuery queries, which are stored in the script $fragments.xq$.
Applied to the set of documents $facts_f.xml$ and $dimension_d.xml$
modeling the target XML data warehouse, these queries finally
produce in output the actual set of fragments, which are stored in a
set of documents $facts_{f_i}.xml$ and $dimension_{d_i}.xml$, with
$i=1, ..., K+1$. These documents represent the final result of the
overall fragmentation process. As fragments, these documents indeed
bear the same schema than the original data warehouse. In
particular, the $(K+1)^{th}$ fragment/document is based on an
additional predicate, named as $ELSE$, which is defined as the
negation of the conjunction of all predicates in $SP$ and it is
necessary to ensure completeness of the fragmentation
(Section~\ref{sec:relational-fragmentation}). In our running
example, $ELSE = \neg (p_1 \wedge  p_2 \wedge  p_3 \wedge p_4)$.

Figure~\ref{fig:frag-q} provides an excerpt from the script
\emph{fragments.xq} that generates fragment $f2$ of
Figure~\ref{fig:frag-doc}. As shown in Figure~\ref{fig:frag-q},
dimension fragments are generated first, one by one, through
selections exploiting the predicate(s) associated to the current
dimension (i.e., the first three queries from
Figure~\ref{fig:frag-q}). Then, fragmentation is derived on facts by
joining the original fact document to the newly-created dimension
fragments (i.e., the last query from Figure~\ref{fig:frag-q}).

\begin{figure}[hbt]
\centering{ \scriptsize{
\begin{tabular}{p{8cm}}
element dimension\{ attribute dim-id\{Customer\}, element Level\{ \\
attribute id \{Customers\},\\
for \$x in document("$dimension_{Customer}.xml$")//Level \\
where \$x//attribute[@id="c\_nation\_key"]/@value="13"] \\
return \$x \} \\
\}\\
element dimension\{ attribute dim-id\{Part\}, element Level\{ \\
attribute id \{Part\},\\
for \$x in document("$dimension_{Part}.xml$")//Level \\
where \$x//attribute[@id="p\_type"]/@value="PBC"] \\
return \$x \} \\
\}\\
element dimension\{ attribute dim-id\{Date\}, element Level\{ \\
attribute id \{Date\},\\
for \$x in document("$dimension_{Date}.xml$")//Level \\
where \$x//attribute[@id="d\_date\_name"]/@value="Sat"] \\
return \$x \} \\
\}\\
element FactDoc \{\\
for \$x in //FactDoc/Fact,\\
    \$y in document("$dimension_{Customer_{f2}}.xml$")//instance,\\
    \$z in document("$dimension_{Part_{f2}}.xml$")//instance,\\
    \$t in document("$dimension_{Date_{f2}}.xml$")//instance\\
where \$x/dimension[@dim-id="Customer"]/@value-id=\$y/@id\\
and \$x/dimension[@dim-id="Part"]/@value-id=\$z/@id\\
and \$x/dimension[@dim-id="Date"]/@value-id=\$t/@id\\
return \$x \\
\}
\end{tabular}}
} \caption{Excerpt from the Script $fragments.xq$ Generating the
Fragment $f2$ of Figure~\ref{fig:frag-doc}} \label{fig:frag-q}
\end{figure}

%%%%%%%%%%%%%%%%%%%%%%%%%%%%%%%%%%%%%%%%%%%%
\section{Experimental Assessment} %% 5
\label{sec:ExperimentalPerformanceStudy}
%%%%%%%%%%%%%%%%%%%%%%%%%%%%%%%%%%%%%%%%%%%%

It has been already demonstrated that derived horizontal
fragmentation is an NP-hard problem~\cite{Boukhalfa08BR}. It follows
that devising a theoretical evaluation of our XML data warehouse
fragmentation technique, even highly significant, would be
particularly hard, although some asymptotic analysis for very simple
cases could be still investigated. Therefore, in this Section we
provide the experimental assessment of our proposed technique, which
gives us a reliable case towards the validation of the effectiveness
and efficiency of the technique.

\subsection{Experimental Settings}

In our experimental assessment, we use \emph{XML Data Warehouse
Benchmark} (XWeB)~\cite{MahboubiD06} as test platform. XWeB is a
benchmark XML data warehouse based on the reference model presented
in Section~\ref{sec:XMLWarehouseReferenceModel}. XWeB also provides
an XQuery-modeled decision-support query-workload that is exploited
to stress the query performance of XML data warehouse query and
processing algorithms running on the benchmark.

XWeB warehouse stores facts related to \emph{Sales} of a typical
retail application scenario, on top of which the following
\textsf{SUM}-based measures are defined: \emph{Amount} and
\emph{Quantity} (of purchased products). Four dimensions complete
the XWeB multidimensional model: (\emph{i}) \emph{Customer}, which
models customers purchasing products; (\emph{ii}) \emph{Supplier},
which models the suppliers furnishing products; (\emph{iii})
\emph{Date}, which models the temporal dimension of the XWeB
warehouse; (\emph{iv}) \emph{Part}, which models the products. Facts
are stored in the document $facts_{Sales}.xml$, whereas dimensions
are stored in the documents
$dimension_{Customer}.xml$, $dimension_{Supplier}.xml$,\\
$dimension_{Date}.xml$ and $dimension_{Part}.xml$, respectively.
XWeB warehouse characteristics are summarized in
Table~\ref{Table:dw-characteristics}.

\begin{table}[hbt]
\begin{center}
\begin{tabular}{|l|l|}
\hline \small{\textbf{Facts}} & \small{\textbf{Maximum Number of Facts}}\\
\hline Sales & $7,000$ \\
\hline
\hline \small{\textbf{Dimensions}} & \small{\textbf{Number of Instances}}\\
\hline Customer & $1,000$ \\
\hline Supplier & $1,000$ \\
\hline Date  & $500$ \\
\hline Part  & $1,000$ \\
\hline
\hline \small{\textbf{Documents}} & \small{\textbf{Size (MB)}}\\
\hline $facts_{Sales}.xml$ & $2.14$\\
\hline $dimension_{Customer}.xml$ & $0.431$\\
\hline $dimension_{Supplier}.xml$ & $0.485$\\
\hline $dimension_{Date}.xml$  & $0.104$\\
\hline $dimension_{Part}.xml$  & $0.388$\\
\hline
\end{tabular}
\caption{XWeB Warehouse Characteristics}
\label{Table:dw-characteristics}
\end{center}
\end{table}

XWeB query-workload is composed by queries that exploit the
warehouse through join and selection operations. In order to obtain
a significant fragmentation, in our experimental assessment we
extend the XWeB workload by adding selection predicates. The
so-obtained workload is available at~\cite{Mahboubi08}.

As regards XML data management aspects, in our experimental
assessment we use the \emph{X-Hive} XML native DBMS~\cite{X-Hive08}
to store and query the data warehouse. As regards the hardware
infrastructure of our experimental framework, we use a \emph{Pentium
Core 2} host at $2$ GHz equipped with $1$ GB RAM and running
\emph{Windows XP}. Finally, our experimental software platform is
written in \emph{Java} and interacts with X-Hive and Weka through
their respective APIs.

\subsection{Comparison Fragmentation Techniques}
\label{sec:FragmentationStrategyComparison}

In our experimental assessment, we compare our proposed
\emph{K}-means-based fragmentation technique (denoted as \emph{KM})
with classical derived horizontal fragmentation techniques, namely
predicate construction (denoted as \emph{PC}) and affinity-based
(denoted as \emph{AB}) primary fragmentation techniques
(Section~\ref{sec:relational-fragmentation}), which we adapt and
specialize to XML data warehouses~\cite{dapd}. In order to compare
even with the baseline instance, we also consider the case in which
no fragmentation is applied (denoted as \emph{NF}).

\subsection{First Experiment: Query Response Time}
\label{sec:QueryResponseTime}

In the first experiment of our experimental campaign, we measure the
query response time needed to evaluate all the queries of the target
query-workload. For what regards \emph{KM}, we arbitrarily fix
$K=8$, which could correspond to the number of hosts of a
conventional cluster of computers. The fragments we obtain are
stored in distinct collections, in order to simulate a reliable
fragment distribution. This well simulates a setting in which each
collection can be considered as stored on a distinct node of the
network on which the data warehouse is distributed, and, moreover,
each collection can be identified, targeted and queried separately.
Overall, this realizes a distributed data warehouse environment
finely. In order to measure the query execution time of the whole
query-workload over the fragmented data warehouse, we first identify
fragments involved by queries thanks to the document
$frag-schema.xml$, and then we execute queries over fragments and
save execution times. To simulate parallel execution, like in a
cluster computer scenario, we consider the maximum execution time.
This provides us with a reliable estimation of the query response
time needed to execute all the queries of the target query-workload
due to a parallel execution.

\begin{figure}[hbt]
\centering \epsfig{file=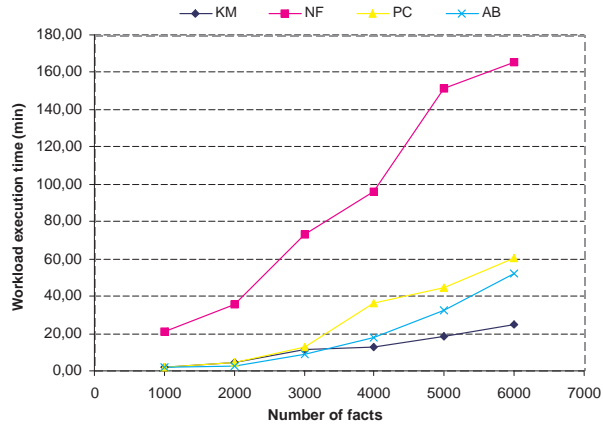, width=9.5cm} \caption{Query
Response Time of Comparison Fragmentation Techniques}
\label{fig:exp11}
\end{figure}

Figure~\ref{fig:exp11} shows the query response time for the target
query-workload with respect to the data warehouse size expressed in
number of facts. The Figure clearly demonstrates that fragmentation
significantly improves query response time, and that \emph{KM}
fragmentation allows us to achieve a better performance than
\emph{PC} and \emph{AB} fragmentation when the warehouse size scales
up. Obviously, \emph{KM} also outperforms \emph{NF}. More precisely,
workload execution time is, on the average, $86.5\%$ faster with
\emph{KM} fragmentation than \emph{PC} fragmentation, and $36.7\%$
faster with \emph{KM} fragmentation than \emph{AB} fragmentation.
Our approach performs better than classical derived horizontal
fragmentation techniques also because the latter techniques
originate much more fragments when compared with ours, i.e. $159$
with \emph{PC} fragmentation, $119$ with \emph{AB} fragmentation and
$9$ with \emph{KM} fragmentation. Hence, when classical
fragmentation techniques are applied, at workload execution time
queries must access a large number of fragments (up to $50$ from our
observations of the actual experiment), which significantly
multiplies both query distribution and result reconstruction costs.
Contrary to this, when the \emph{KM} fragmentation technique is
applied, the number of accessed fragments is much lower (typically
$2$ fragments in the actual experiment).

\subsection{Second Experiment: Fragmentation Cost}
\label{sec:FragmentationOverhead}

In the second experiment of our experimental campaign, we compare
the \emph{PC}, \emph{AB} and \emph{KM} ($K=8$) fragmentation
strategies in terms of fragmentation costs, i.e. we investigate the
execution time of proper fragmentation algorithms. Before going into
details, we focus the attention on the complexity of fragmentation
algorithms. Let $|SP|$ denote the cardinality of the selection
predicate set $SP$. It follows than the algorithm complexities for
the comparison fragmentation techniques are the following:
$O(2^{|SP|})$ for \emph{PC}, $O(|SP|^2)$ for \emph{AB}, and
$O(|SP|)$ for \emph{KM} fragmentation technique. Therefore, on a
pure theoretical plane, our proposed XML data warehouse
fragmentation technique exposes a complexity lower that those of
comparison approaches.

Indeed, despite theoretical issues, when algorithms' performance is
evaluated, it is necessary to find a fair trade-off between
effective gain and computational overheads. Therefore, it is
mandatory to develop a reliable experimental evaluation. In this
respect, Table~\ref{Table:exp12} summarizes the results we obtain
for an arbitrarily-fixed data warehouse size equal to $3,000$ facts.
Obtained results clearly show that \emph{KM} fragmentation technique
outperforms both \emph{PC} and \emph{AB} fragmentation techniques.

It should be noted that our results are not fully-in-line with
above-introduced algorithms' complexities, as in our experimental
assessment we include the time required by constructing fragments in
the overall evaluation of computational overheads of algorithms.
Hence, since \emph{PC} and \emph{AB} fragmentation techniques
originate a large number of fragments, building such fragments
requires a large number of costly join operations accordingly, thus
leading to long running times. An immediate conclusion coming from
this experimental evidence states that, while \emph{PC} and
\emph{AB} fragmentation techniques are likely to run in an offline
manner, \emph{KM} fragmentation technique could on the other hand be
envisaged to run in an online manner, thus turning to be perfectly
suitable to OLAP applications.

\begin{table}[hbt]
\begin{center}
\begin{tabular}{|l|c|c|c|}
\hline
& \small{\textbf{\emph{PC}}} & \small{\textbf{\emph{AB}}} & \small{\textbf{\emph{KM}}} \\
\hline
\small{\textbf{Execution Time (h)}} & \small{$16.8$} & \small{$11.9$} & \small{$0.25$} \\
\hline
\end{tabular}
\caption{Fragmentation Cost of Comparison Fragmentation Techniques}
\label{Table:exp12}
\end{center}
\end{table}

\subsection{Third Experiment: Influence of the Number of Clusters}
\label{sec:InfluenceOfNumberOfClusters}

\begin{figure}[hbt]
\centering \epsfig{file=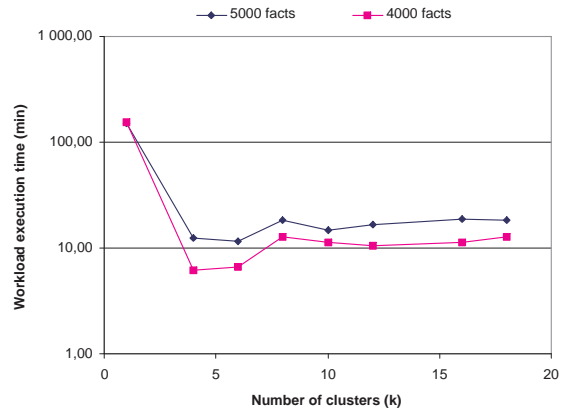, width=8cm} \caption{Influence of
the Number of Clusters for the \emph{KM} Fragmentation Technique}
\label{fig:exp2}
\end{figure}

In the third experiment of our experimental campaign, we fix the
data warehouse size to $4,000$ and $5,000$ facts, respectively, and
vary the parameter $K$ of the \emph{KM} fragmentation technique in
order to observe the influence of the number of clusters on the
workload response time. Figure~\ref{fig:exp2} confirms that
performance improves quickly when fragmentation is applied, but it
tends to degrade when the number of fragments increases, according
to the discussion provided in Section~\ref{sec:QueryResponseTime}.
Furthermore, results depicted in Figure~\ref{fig:exp2} suggest to us
that the optimal number of clusters for our benchmark data warehouse
and related query-workload lies between $4$ and $6$, which allows us
to conclude that over-fragmentation (i.e., generating an excessive
number of fragments) must be detected and avoided in distributed
data warehouses (note that, in Figure~\ref{fig:exp2}, $K=1$
corresponds to the \emph{NF} experimental setting, i.e. one fragment
corresponding to the original data warehouse).

%%%%%%%%%%%%%%%%%%%%%%%%%%%%%%%%%%%%%%%%%%%%%%
\section{Conclusions and Future Work} %% 6
\label{sec:Conclusion}
%%%%%%%%%%%%%%%%%%%%%%%%%%%%%%%%%%%%%%%%%%%%%%

In this paper, we have introduced an approach for fragmenting XML
data warehouses that is based on Data Mining, and, more precisely,
on \emph{K}-means clustering algorithm. Classical derived horizontal
fragmentation strategies run automatically, and output an
unpredictable number of fragments, which is indeed nonetheless
crucial to keep under control in realistic distributed data
warehouses. By contrary, our proposed fragmentation approach allows
us to fully master the number of fragments through the parameter $K$
of \emph{K}-means algorithm.

In order to validate the effectiveness and the efficiency of our
proposal, we have compared our fragmentation strategy to meaningful
adaptations of the two prevalent fragmentation methods for
relational data warehouses, i.e. the \emph{PC} and \emph{AB}
fragmentation techniques, to the specialized context of XML data
warehouses. Obtained experimental results show that our approach
significantly outperforms both comparison techniques (along with the
baseline case in which no fragmentation is applied) under several
perspective of experimental analysis.

Upon the fragmentation results above, future work is focused to the
problem of effectively and efficiently distributing XML data
warehouses on data grids. This issue raises several challenges that
include decomposing a global query posed to the \emph{grid-enabled
XML data warehouse} into a set of sub-queries to be sent to the
correct grid nodes, and meaningifully reconstructing the global
result from intermediate sub-query results. In this direction,
properly indexing the distributed data warehouse in order to
guarantee good performance seems to be a critical aspect.

Finally, in a continuous effort towards minimizing data warehouse
administration functions and aiming at auto-administrative
systems~\cite{adbis06,innovations07}, we plan to make \emph{dynamic}
our Data-Mining-based fragmentation approach. Here, the main idea
consists in performing \emph{incremental fragmentation} as long as
the target data warehouse is refreshed (e.g., during maintenance
operations). This could be achieved by exploiting an
\emph{incremental} variant of \emph{K}-means clustering
algorithm~\cite{RT05}.

\NUMBIB

\end{document}